\documentclass[sigconf]{acmart}
\usepackage{booktabs} % For formal tables

\usepackage{natbib}

\acmConference[]{} 
\acmYear{}
\copyrightyear{}

\renewcommand\footnotetextcopyrightpermission[1]{} % removes footnote with conference information in first column
\begin{document}
\title{Energy-Efficient Time-Domain Vector-by-Matrix Multiplier for Neurocomputing and Beyond}

\author{Mohammad Bavandpour, Mohammad Reza Mahmoodi, and Dmitri B. Strukov}
\affiliation{%
  \institution{UC Santa Barbara, Department of Electrical and Computer Engineering, Santa Barbara, CA, 93106, U.S.A}
}
\email{{mbavandpour, mrmahmoodi, strukov}@ece.ucsb.edu}

\begin{abstract}

We propose an extremely energy-efficient mixed-signal approach for performing vector-by-matrix multiplication in a time domain. In such implementation, multi-bit values of the input and output vector elements are represented with time-encoded digital signals, while multi-bit matrix weights are realized with current sources, e.g. transistors biased in subthreshold regime. With our approach, multipliers can be chained together to implement large-scale circuits completely in a time domain. Multiplier operation does not rely on energy-taxing static currents, which are typical for peripheral and input/output conversion circuits of the conventional mixed-signal implementations. As a case study, we have designed a multilayer perceptron, based on two layers of $10\times10$ four-quadrant vector-by-matrix multipliers, in 55-nm process with embedded NOR flash memory technology, which allows for compact implementation of adjustable current sources. Our analysis, based on memory cell measurements, shows that at high computing speed the drain-induced barrier lowering is a major factor limiting multiplier precision to $\sim 6$ bit. Post-layout estimates for a conservative 6-bit digital input/output $N\times N$ multiplier designed in 55 nm process, including I/O circuitry for converting between digital and time domain representations, show $\sim7$ fJ/Op for $N>200$, which can be further lowered well below 1 fJ/Op for more optimal and aggressive design. 

\end{abstract}

\maketitle

\section{Introduction}
\par The need for computing power is steadily increasing across all computing domains, and has been rapidly accelerating recently in part due to the emergence of machine learning, bioinformatics, and internet-of-things applications. With conventional device technology scaling slowing down and many traditional approaches for improving computing power reaching their limits, the increasing focus now is on heterogeneous computing with application specific circuits and specialized processors \cite{tpu, asic}. 

\par Vector-by-matrix multiplication (VMM) is one of the most common operations in many computing applications and, therefore, the development of its efficient hardware is of the utmost importance. (We will use VMM acronym to refer to both vector-by-matrix multiplication and vector-by-matrix multiplier.)  Indeed, VMM is a core computation in virtually any neuromorphic network \cite{deep1} and many signal processing algorithms \cite{sigprocess2,fg3}. For example, VMM is by far the most critical operation of deep-learning convolutional classifiers \cite{envision,memdpe2}. Theoretical studies showed that sub-8-bit precision is typically sufficient for the inference computation \cite{quant}, which is why the most recent high-performance graphics processors support 8-bit fixed-point arithmetics \cite{p40}. The low precision is also adequate for lossy compression algorithms, e.g. those based on discrete cosine transform, which heavily rely on VMM operations \cite{lossydct}.  In another recent study, dense low-precision VMM accelerators were proposed to dramatically improve performance and energy efficiency of linear sparse system solvers, which may take months of processing time when using conventional modern supercomputers \cite{ipek}. In such solvers, the solution is first approximated with dense low-precision VMM accelerators, and then iteratively improved by using traditional digital processors. 

\par The most promising implementations of low to medium precision VMMs are arguably based on analog and mixed-signal circuits \cite{indiveri}. In a current-mode implementation, the multi-bit inputs are encoded as analog voltages/currents, or digital voltage pulses \cite{reram}, which are applied to one set of (e.g., row) electrodes of the array with adjustable conductance cross-point devices, such as memristors \cite{memdpe2, reram} or floating-gate memories \cite{fg1,fg3, sstVMM}, while VMM outputs are represented by the currents flowing into the column electrodes. The main drawback of this approach is  energy-hungry and area-demanding peripheral circuits, which, e.g. rely on large static currents to provide accurate virtual ground for the memristor implementation.

\begin{figure*}[htbp]
\normalsize
\centering
\includegraphics[width=16cm]{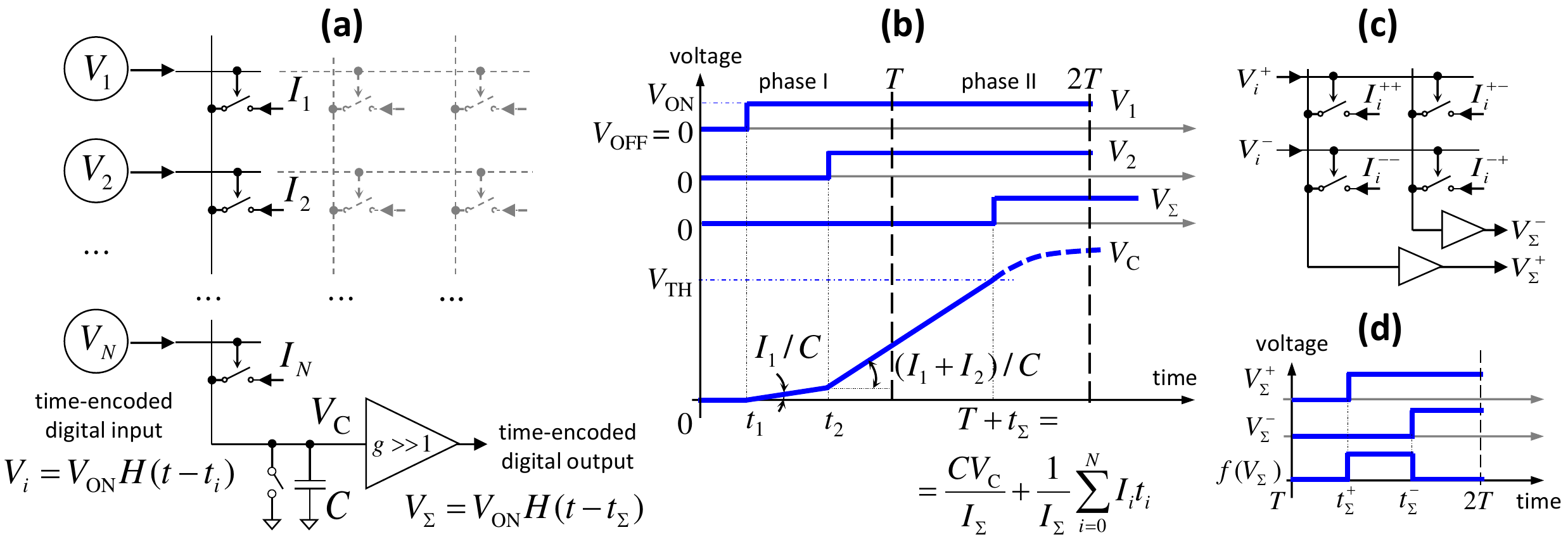}
\caption{\small{The main idea of time-domain vector-by-matrix multiplier: (a) Circuit and (b) timing diagrams explaining the operation for single-quadrant multiplier, assuming $V_{\textrm{OFF}}=0$ so that the inputs and outputs are conveniently described with Heaviside function. Note that panel a does not show bias input $I_0$. (c) Four-quadrant multiplier circuit diagram, showing for clarity only one matrix weight, implemented with four current sources $I_i^{++}$, $I_i^{--}$, $I_i^{-+}$, and $I_i^{+-}$. (d) The timing diagram showing example of the positive output of the four-quadrant multiplier. The bottom diagram for $f(V_\Sigma)$ corresponds to the case-study circuit implementation (Fig.~\ref{mapping}).}}
\label{general}
\end{figure*} 

\par In principle, the switched-capacitor approach does not have this deficiency since the computation is performed by only moving charges between capacitors~\cite{swcap}. Unfortunately, the benefits of such potentially extremely energy-efficient computation are largely negated at the interface, since reading the output, or cascading multiple VMMs, require energy-hungry ADC and/or analog buffers. Perhaps even more serious drawback, significantly impacting the density (and as a result other performance metrics), is a lack of adjustable capacitors. Instead, each cross-point element is typically implemented with the set of fixed binary-weighted capacitors coupled to a number of switches (transistors), which are controlled by the digitally-stored weight.

\par In this paper we propose to perform vector-by-matrix multiplication in a time-domain, combining configurability and high density of the current-mode implementation with energy-efficiency of the switch-capacitor VMM, however, avoiding costly I/O conversion of the latter. Our approach draws inspiration from prior work on time-domain computing \cite{time1,time2,time4,time5,timebnn}, but different in several important aspects.  The main difference with respect to  Refs. \cite{timebnn, time4, time5} is that our approach allows for precise four-quadrant VMM using analog input and weights. Unlike the work presented in Ref.~\cite{time1}, there is no weight-dependent scaling factor in the  time-encoded outputs, making it possible to chain multipliers together to implement functional large-scale circuits completely in a time domain. Inputs were encoded in the duration of the digital pulses in recent work \cite{reram}. However, that approach shares many similar problems of current-mode VMMs, most importantly of costly I/O conversion circuits. Moreover, due to resistive synapses considered in Ref. \cite{reram}, the output line voltage must be pinned for accurate integration which results in large static power consumption. Finally, in this paper, we are presenting preliminary, post-layout performance results for a simple though representative circuit. Our estimates are based on the layout of the circuit in 55 nm process with embedded NOR flash memory floating gate technology, which is very promising for the proposed time-domain computing.

\section{Time-Domain Vector-by-Matrix Multiplier}
\subsection{Single-Quadrant Dot-Product Operation}
\par Let us first focus on implementing $N$-element time-domain dot-product (weighted-sum) operation of the form
\begin{equation}
\label{0}
  y = \frac{1}{N w_{\textrm{max}}}\sum_{i=1}^{N}w_i x_i,
\end{equation}
with non-negative inputs $x_i$ and output $y$, and with weights $w_i$ in a range of [0, $w_\textrm{max}$]. Note that for convenience of chaining multiple operations, the output range is similar to that of the input due to normalization. 

\par  The $i$-th input and the dot-product output are time-encoded, respectively, with digital voltages $V_i$ and $V_\Sigma$, such that:
\begin{equation}
\label{1}
  V_i=\left\{ 
  \begin{array}{l l}
   V_{\textrm{OFF}} & 0\leq t<t_i\leq T\\
   V_{\textrm{ON}} & t_i\leq t<2T \\
  \end{array} \right.
\end{equation}
\begin{equation}
\label{2}
  V_\Sigma=\left\{ 
  \begin{array}{l l}
   V_{\textrm{OFF}} & T\leq t<T+t_\Sigma \leq 2T\\
   V_{\textrm{ON}} & T+ t_\Sigma \leq t<3T \\
  \end{array} \right.
\end{equation}
Here, value of $T-t_i \propto x_i$ represents a multi-bit (analog) input, which is defined within a time window $[0, T]$, while the value $T-t_\Sigma \propto y$ represents a multi-bit (analog) output observed within a time window $[T, 2T]$. With such definitions, the maximum values for the inputs and the output are always equal to $T$, their minimum values are 0, while $V_i \equiv V_\textrm{ON}$ for $T\leq t<2T$ and $V_\Sigma \equiv V_\textrm{ON}$ for $2T\leq t<3T$.
\par The time-encoded voltage inputs are applied to the array's row electrodes, which connect to the control input of the adjustable configurable current sources, i.e. gate terminals of transistors (Fig. \ref{general}a). $V_{\textrm{OFF}}$ input is assumed to  turn off the current source (transistor). On the other hand, application of $V_{\textrm{ON}}$ voltage  initiates the constant current $0 \leq I_i \leq I_\textrm{max}$, specific to the programmed value of the $i$-th current source, to flow into the column electrode. This current will charge the capacitor $C$ which is comprised by the capacitance of the column (drain) line and and intentionally added external capacitor. When the capacitor voltage $V_\textrm{C}$ reaches threshold $V_{\textrm{TH}}$, the output of the digital buffer will switch from $V_{\textrm{OFF}}$ to $V_{\textrm{ON}}$ at time $T+t_\Sigma$, which is effectively the time-encoded output of the dot-product operation.

\par Assuming for simplicity $V_\textrm{C}(t=0)=V_\textrm{OFF}=0$ and negligible dependence of the currents injected by current sources on $V_\textrm{C}$ (e.g., achieved by biasing cross-point transistor in sub-threshold regime) the charging dynamics of the capacitor is given by the dot product of currents $I_i H(t-t_i)$, where $H$ is Heaviside function, with their corresponding time intervals $t-t_i$, i.e
\begin{equation}
\label{3}
  CV_\textrm{C}(t)=\sum_{i=0}^{N}I_iH(t-t_i)(t-t_i), \indent 0 \leq t \leq 2T .
\end{equation}

Before deriving the expression for the output $T-t_\Sigma$, let us note two important conditions imposed by our assumptions for the minimum and maximum values of the output. First, the additional (weight-dependent) bias current $I_0$ is added to the sum in Eq. \ref{3}, with its turn-on time always $t_0=0$, to make output  $T-t_\Sigma$ equal to 0 for the smallest possible value of dot product, represented by $t_{1,2,...,N} = T$ and $I_{1,2,...,N} = 0$. Secondly, the largest possible output $T-t_\Sigma = T$, which corresponds to the largest values of inputs $T-t_{1,2,...,N}=T$ and current sources $I_{1,2,...N}=I_\textrm{max}$, is ensured by
\begin{equation}
\label{7}
  I_{\textrm{max}}=\frac{CV_{\textrm{TH}}}{NT}.
\end{equation}
Using $V = V_{\textrm{TH}}$ in Eq. \ref{3} and noting that $H$ is always 1 for $t_\Sigma \geq 0$, the relation between time-encoded output and the inputs is described similarly to Eq. \ref{0}
%, i.e.  
%\begin{equation}
%\label{4}
%  T-t_\Sigma = \frac{1}{N w_{\textrm{max}}}\sum_{i=1}^{N}w_i (T-t_i),
%\end{equation}
if we assume that currents $I_i$ are 
\begin{equation}
\label{4}
  I_i=I_{\textrm{max}} \frac{CV_{\textrm{TH}}w_i}{2 CV_{\textrm{TH}}w_{\textrm{max}}-I_{\textrm{max}}T \sum_{i=1}^{N}w_i}, 
  i=1,2,...N,
\end{equation}
and the bias current is
\begin{equation}
\begin{split}
\label{5}
  I_0=\frac{1}{2}\Big(\frac{CV_{\textrm{TH}}}{T}-\sum_{i=1}^{N}I_i\Big)
  \equiv \frac{1}{2}\Big(N I_{\textrm{max}}-\sum_{i=1}^{N}I_i\Big).
\end{split}
\end{equation}

\subsection{Four-Quadrant Multiplier}
\par  The extension of the proposed time-domain dot-product computation to time-domain VMM is achieved by utilizing the array of current source elements and performing multiple weighted-sum operations in parallel (Fig.~\ref{general}a). The implementation of four-quadrant multiplier, in which input, outputs, and weights can be of both polarities, is shown in Figure \ref{general}c. In such differential style implementation, dedicated wires are utilized for time-encoded  positive and negative inputs / outputs, while each weight is represented by four current sources. The magnitude of the computed output is encoded by the delay, just as it was discussed for single-quadrant dot-product operation, while the output's sign is explicitly defined by the corresponding wire of a pair. To multiply input by the positive weight, $I_i^{++}$ and $I_i^{--}$ are set to the same value according to Eq. 7, with the other current source pair set to $I_i^{+-} =  I_i^{-+} = 0$, while it is the opposite for the multiplication by the negative weights. Note that with such implementation, in the most general case, the input / output values are represented by time-encoded voltage pulses on both positive and negative wires of the pair, with, e.g., $t_\Sigma^+-t_\Sigma^-<0$ corresponding to the positive output (Fig.~\ref{general}d), and negative (or zero, when $t_\Sigma^+-t_\Sigma^-=0$)  output otherwise.  

\par The output for the proposed time-domain VMM is always computed within fixed $2T$-long window. Its maximum and minimum values are independent of a particular set of utilized weights and always correspond to $T$ and $2T$, which is different from prior proposals \cite{time1}. This property of our approach is convenient for implementing large-scale circuits completely in a time-domain. For example, VMM outputs can be supplied directly to another VMM block or some other time-domain circuitry, e.g. implementing Race Logic \cite{racelogic}.

\begin{figure}[t]
\normalsize
\centering
\includegraphics[width=8cm]{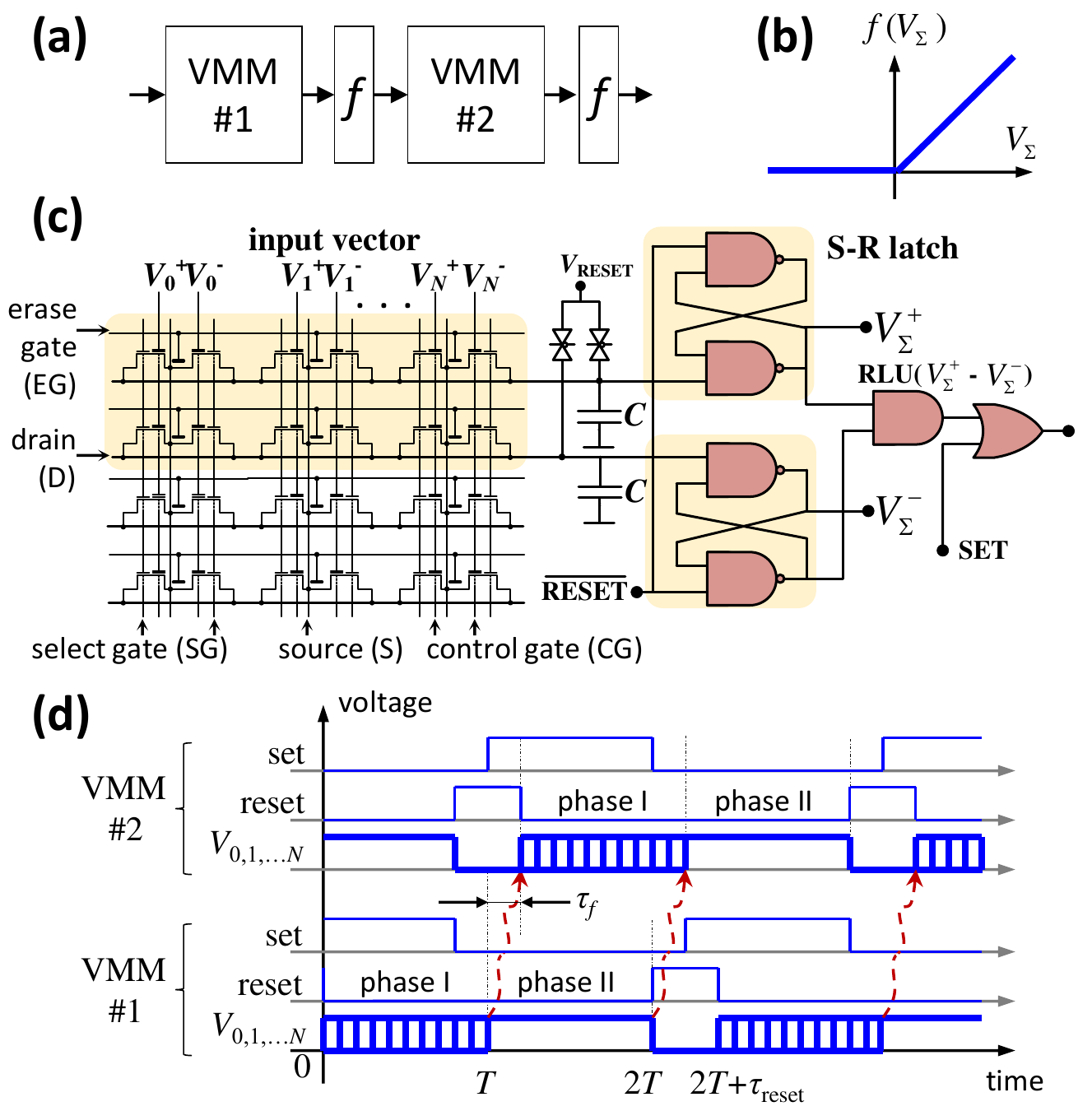}
\caption{\small{Two-layer perceptron network: (a) Block diagram of the considered circuit with (b) specific rectify-linear function. (c) Implementation details of VMM and rectify-linear circuits. Panel (c) shows only peripheral circuity required for four-quadrant dot-product operation, which involves two rows of adjustable current sources (floating gate transistors) and two S-R latches (highlighted with yellow background). (d) Timing diagram for pipelined operation (shown schematically). Here $\tau_f$ is a combined propagation delay of S-R latch and rectify-linear circuit. The dashed red arrows show the flow of information between two VMMs. Note that the inputs to VMMs are always high during the phase II.}}
\label{mapping}
\end{figure}

\section{Case Study: Perceptron Network}

\subsection{Design Methodology}

\par The design methodology for implementing larger circuits is demonstrated on the example of specific circuit comprised of two VMMs and a nonlinear (``rectify-linear'') function block, which is representative of multi-layer perceptron network (Fig. \ref{mapping}a, b). Figure~\ref{mapping}c shows gate-level implementation of VMM and rectify-linear circuits, which is suitable for pipelined operation. Specifically, the thresholding  (digital buffer) is implemented with S-R latch. The rectify-linear functionality is realized with just one AND gate which takes input from two latches that are serving a differential pair. The AND gate will generate a voltage pulse with $t_\Sigma^--t_\Sigma^+$ duration for positive outputs from the first VMM (see, e.g., Fig.~\ref{general}d) or have zero voltage for negative ones. 

\par It is important to note that in the considered implementation of rectify-linear function, the inputs to the second VMM are encoded not in the rising edge time of the voltage pulse, i.e. $T-t_i$ , but rather in its duration. Specifically, in this case the input is encoded by $t_i$-long voltage pulse in the phase I (i.e. $[0, T]$ time interval), and always $T$-long voltage pulse during the phase II ($[T,2T]$ time interval). Such pulse-duration encoding is more general case of a scheme presented in Section 2.1. Indeed, in our approach, each product term in dot-product computation is contributed by the total charge injected to an output capacitor by one current source, which in turn is proportional to the total time that current source is on. In the scheme discussed in Section 2.1, voltage pulse always ends at $2T$ and thus encoding in the rising edge time of a pulse is equivalent to the encoding in the pulse duration. 

\par Additional pass gates, one per each output line, are  controlled by RESET signals (Fig. \ref{mapping}c) and are used to pre-charge output capacitor before starting new computation.  Controlled by SET signal, the output OR gate is used to decouple computations in two adjacent VMMs. Specifically, the OR gate and SET signal allow to generate phase II's $T$-long pulses applied to the second VMM and, at the same time, pre-charge and start new phase I computation in the first VMM. Using appropriate periodic synchronous SET and RESET signals, pipelined operation with period $2T+\tau_\textrm{reset}$ is established, where $\tau_\textrm{reset}$ is a time needed to pre-charge output capacitor (Fig. \ref{mapping}d).  

\par Though the slope of activation function is equal to one in the considered implementation, it can be easily controlled by appropriate scaling of VMM weights (either in one or both VMMs). Also, because of strictly positive inputs, in principle, only two-quadrant multiplier is needed for the second layer, which is easily implemented by removing all input wires carrying negative values in the four-quadrant design (Fig. \ref{general}c).

\begin{figure}[t]
\normalsize
\centering
\includegraphics[width=8cm]{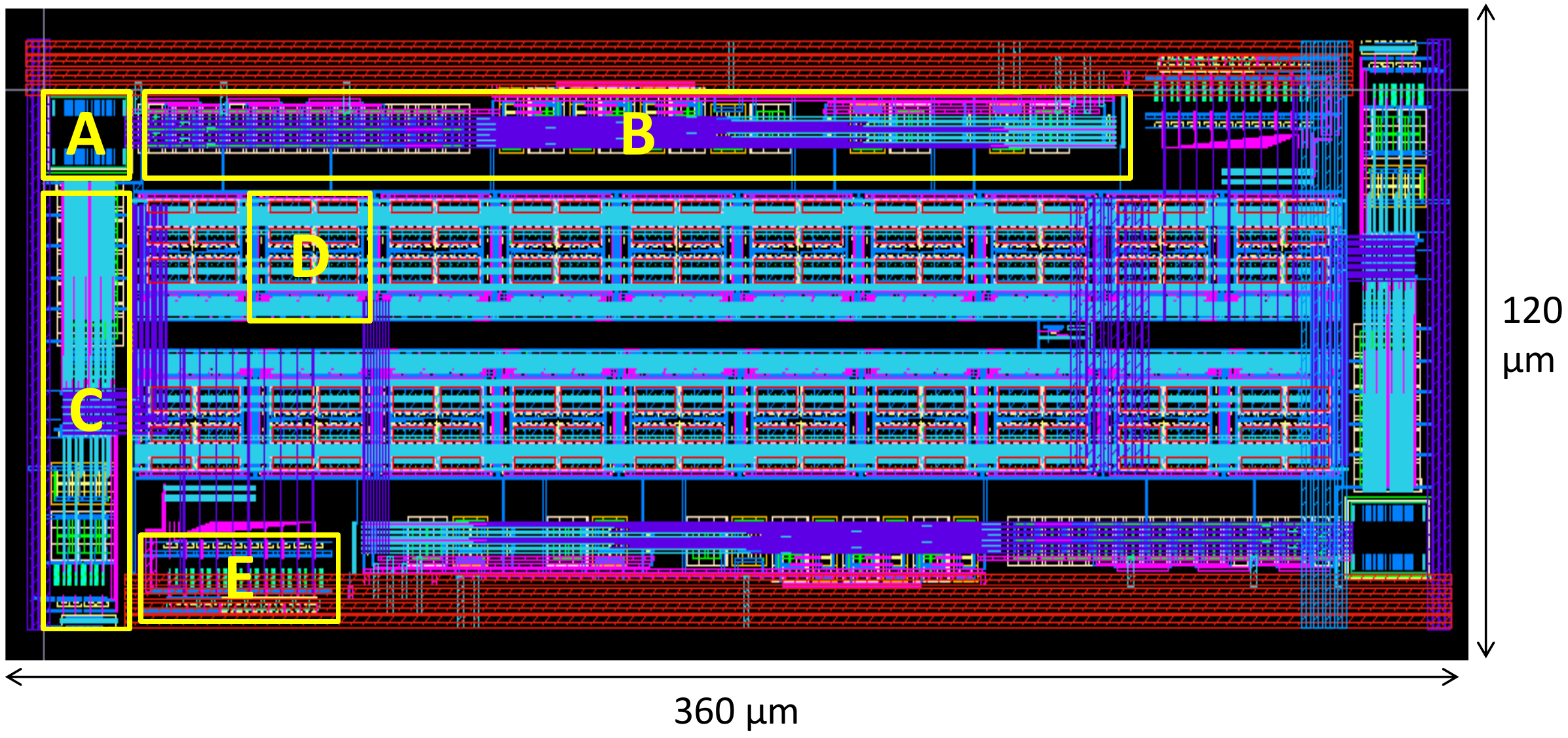}
\caption{\small{Layout of the implemented $10 \times 10 \times 10$ network. Here, label A denotes $10\times20$ supercell array, B/C shows column/row program and erase circuitry, D is one ``neuron'' block, which includes conservative 0.4 pF MOSCAP output capacitor, S-R latch based on ($W=120$ nm, $L=900$ nm) transistors, pass gates, and rectify-linear / pipelining circuit, and E is output multiplexer. All clock and control signals are generated externally.}}
\label{circuit}
\end{figure}

\subsection{Embedded NOR Flash Memory Implementation}
\par We have designed two-layer perceptron network, based on two $10\times10$ four-quadrant multipliers, in 55 nm CMOS process with modified embedded ESF3 NOR flash memory technology \cite{sst}. 
In such technology, erase gate lines in the memory cell matrix  were rerouted (Fig. \ref{mapping}c) to enable precise individual tuning of the floating gate (FG) cells' conductances. The details on the redesigned structure, static and dynamic $I$-$V$ characteristics, analog retention, and noise of the floating gate transistors, as well as results of high precision tuning experiments can be found in Ref. \cite{sstVMM}. 

\par The network, which features 10 inputs and 10 hidden-layer / output neurons, is implemented with two identical $10\times20$ arrays of supercells, CMOS circuits for the pipelined VMM operation and rectify-linear transfer function as described in previous subsection, as well as CMOS circuitry for programming and  erasure of the FG cells (Fig. \ref{circuit}). During operation, all FG transistors are biased in subthreshold regime. The input voltages are applied to control gate lines, while the output currents are supplied by the drain lines. Because FG transistors are N-type, the output lines are not discharged to the ground with RESET signal, but rather charged to $V_{\textrm{RESET}} = V_{\textrm{TH}}+\Delta V_{\textrm{D}}$, i.e.  $\Delta V_{\textrm{D}}$ above the threshold voltage $V_{\textrm{TH}}$ of S-R latch. In this case, VMM operation is  performed by sinking currents via adjustable current sources based on FG memory cells.

\section{Design Tradeoffs and Performance Estimates} 

\par Obviously, understanding the true potentials of the proposed time-domain computing would require choosing optimal operating conditions and careful tuning of the CMOS circuit parameters. Furthermore, the optimal solution will differ depending on the specific optimization goals and input constrains such as VMM size and precision of operation. Here, we discuss important tradeoffs and factors at play in the optimization process, focusing specifically on the computing precision. We then provide preliminary estimates for area, performance, and energy efficiency. 

\begin{figure}[t]
\normalsize
\centering
\includegraphics[width=8cm]{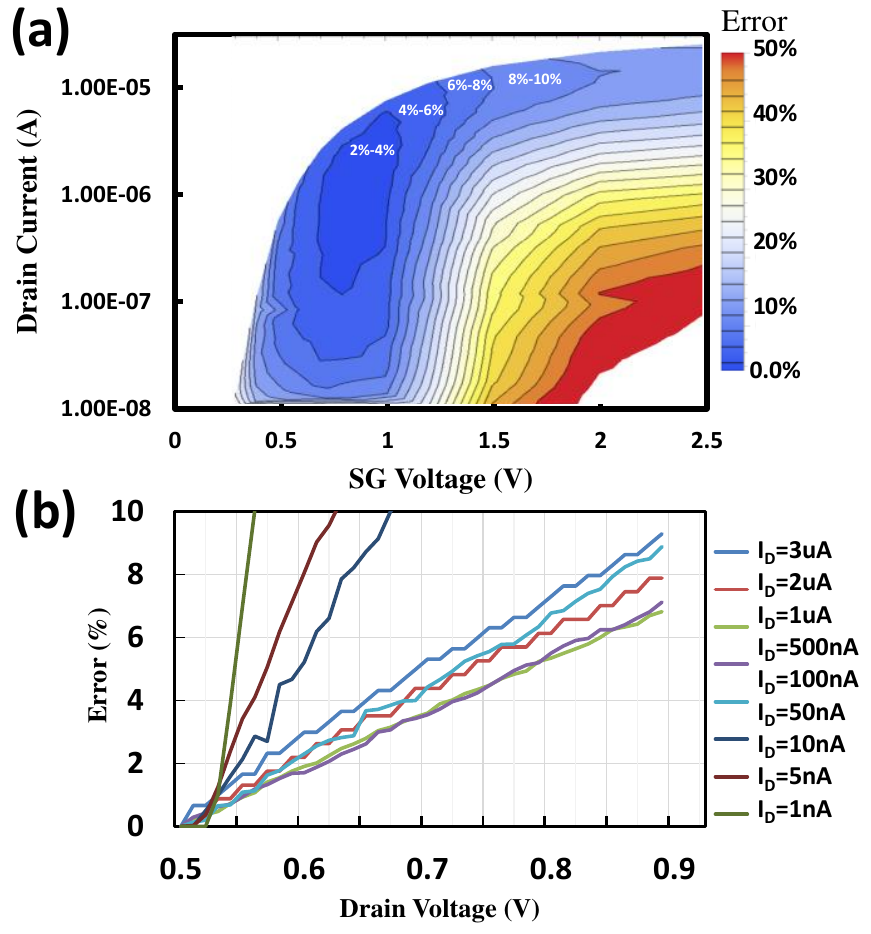}
\caption{\small{Relative output error shown as (a) function of maximum drain current $I_{\textrm{max}}$ and select gate voltage, assuming drain voltage variation $\Delta V_\textrm{D} = 0.2$ V, and (b) function of drain voltage at $V_{\textrm{SG}}=0.8$ V. The error in both panels was calculated by tuning cells to different memory states and  experimentally measuring changes in subthreshold drain current. In all cases, $V_{\textrm{CG}}=$1.2 V and  $V_{\textrm{S}}=V_{\textrm{EG}}=0$ V.}}
\label{bldep}
\end{figure}

\subsection{Precision} There are number of factors which may limit computing precision. The weight precision is affected by the tuning accuracy, drift of analog memory state, and drain current fluctuations due to intrinsic cell's noise. These issues can be further aggravated by variations in ambient temperature. Earlier, it has been shown that, at least in small-scale current-mode VMM circuits based on similar 55-nm flash technology, even without any optimization, all these factors combined may allow up to 8 bit effective precision for majority of the weights \cite{sstVMM}. We expect that just like for current-mode VMM circuits, the temperature sensitivity will be improved due to differential design and utilization of higher drain currents, which is generally desired for optimal design.

Our analysis shows that for the considered memory technology, the main challenge for VMM precision is non-negligible dependence of FG transistor subthreshold currents on the drain voltage (Fig. \ref{bldep}), due to the drain-induced barrier lowering (DIBL). To cope with this problem, we minimized the relative output error $\textrm{Error} = |I_\textrm{max}(V_\textrm{RESET})-I_\textrm{max}(V_\textrm{RESET}-\Delta V_\textrm{D})|/I_\textrm{max}(V_\textrm{RESET})$.  In general, $\textrm{Error}$ depends on the pre-charged drain voltage $V_\textrm{RESET}$, the voltage swing on the drain line $\Delta V_{\textrm{D}}= V_\textrm{RESET}-V_\textrm{TH}$, control and select gate voltages, and the maximum drain current utilized for weight encoding. In our initial study, we assumed that $V_\textrm{RESET} = 0.7$ V, which cannot be too small, due to $I_{\textrm{D}}\propto 1-\textrm{exp}(-V_\textrm{D}/V_\textrm{T})$ in subthreshold regime (with $V_\textrm{S}=0$ V), but otherwise has weak impact on  $\textrm{Error}$. Furthermore, for simplicity, we assumed that $\Delta V_{\textrm{D}} = 0.2$ V, which cannot be too low because of static and short-circuit leakages in CMOS gates (see below), and that $V_{\textrm{CG}} = 1.2$ V, which is a standard CMOS logic voltage in 55 nm process. We found that the drain current is especially sensitive to select gate voltages with the distinct optimum at $V_\textrm{SG} \sim 0.8$ V  (Fig. \ref{bldep}a). This is apparently due to shorter effective channel length for higher $V_\textrm{SG}$, and hence more severe DIBL, and voltage divider effect at lower $V_\textrm{SG}$. Furthermore, the drain dependency is the smallest at higher drain currents $I_\textrm{max} \sim$ 1 $\mu$A (Fig. \ref{bldep}a), though naturally bounded by the upper limit of the subthreshold conduction (Fig. \ref{bldep}b). At such optimal conditions, $\textrm{Error}$ could be less than  $2\%$ (Fig. \ref{bldep}a), which is ensuring at least 5 bits of computing precision.   

\par Similarly to all FG-based analog computing, majority of the process variations, a typical concern for any analog or mixed-signal circuits, can be efficiently compensated by adjusting currents of FG devices, provided that such variations can be properly characterized. This, e.g., includes $V_\textrm{TH}$ mismatches (which can be up to 20 mV rms for the implemented S-R latch according to our Monte Carlo simulations). The only input dependent error, which cannot be easily compensated, is due to the variations in the slope of the transfer characteristics of S-R latch gates. However, this does not seem to be a serious issue for our design, given that the threshold for the drain voltage is always crossed at the same voltage slew rate. Also, note that variations in sub-threshold slope are not important for our circuit because of digital input voltages.

\par VMM precision can be also impacted by the factors similar to those of switch-capacitor approach, including leakages via OFF-state FG devices, channel charge injection from pass transistor at the RESET phase, and capacitive coupling of the drain lines. Fortunately, the dominating coupling between D and CG lines is input-independent and can be again compensated by adjusting the weights. 

\begin{figure}[t]
\normalsize
\centering
\includegraphics[width=8cm]{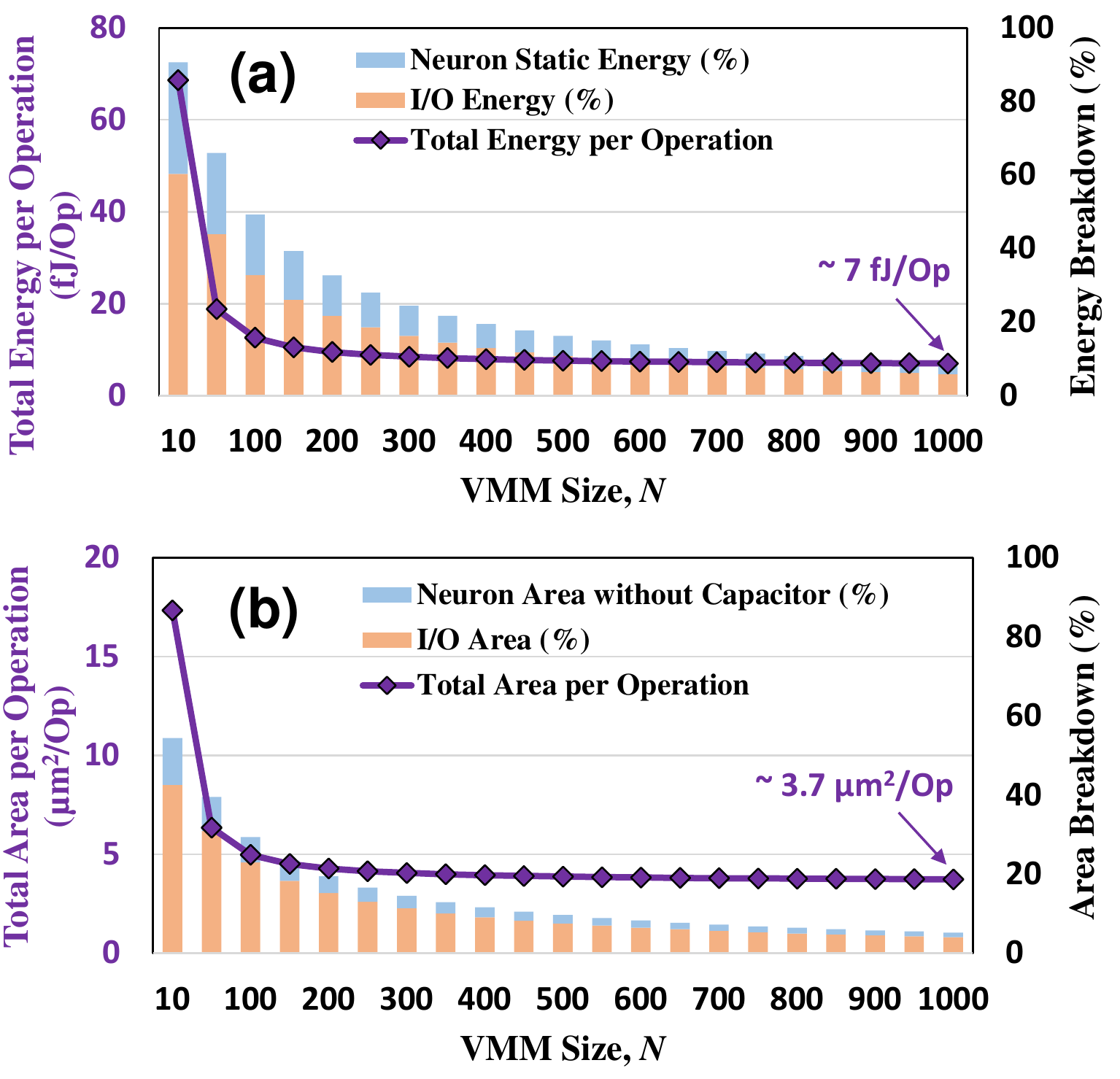}
\caption{\small{(a) Energy and (b) area per operation and their breakdowns for a 6-bit digital-input digital-output time-domain VMM as a function of its size for the conservative case with $C \approx 200  C_\textrm{drain}$. In particular, the bar chart shows contribution of I/O circuitry for converting between digital and time-domain representation (red column), neuron circuitry excluding capactor (cyan column), and memory cell array and external capacitor (remaining white).  Erasure/programming circuitry was not included in the area estimates because it can be shared among multiple VMMs. $N$ is incremented by 50, except for the first bar, which shows data for $10\times10$ VMM.}}
\label{efficiency}
\end{figure}

\subsection{Latency, Energy, and Area } 

\par For a given VMM size $N$, the latency (i.e. $2T$) is decreased by using higher $I_\textrm{max}$ and/or reducing $C$. The obvious limitation to both are degradation in precision, as discussed in previous section, and also intrinsic parasitics of the array, most importantly drain line capacitance $C_\textrm{drain}$. Our post-layout analysis shows that for the implemented circuit and optimal $I_\textrm{max}$, the latency per single bit of computing precision is $2T_0\leq 1$ ns, and roughly  $2T_02^p$ for higher precision $p$, e.g., $\sim 100$ ns for 6-bit VMM. These estimates are somewhat pessimistic because of conservative choice of $C \approx 200 C_\textrm{drain} = 0.04$ pF per input, which ensures <0.5\% votlage drop on the drain line due to capacitive coupling. (Note that in the implemented VMM circuit the intrinsic delay does not scale with array size because of the utilized bootstrapping technique.) 

\par The energy per operation is contributed by the dynamic component of charging/discharging of control gate and drain lines as well as external output capacitors, and the static component, including vdd-to-ground and short-circuit leakages in the digital logic. Naturally, CMOS leakage currents are suppressed exponentially by increasing drain voltage swing, and can be further reduced by lowering CMOS transistor currents (i.e. increasing length to width ratio), while still keeping propagation delay $\tau_\textrm{f}$ negligible as compared to $T$. The increase in the drain swing, however, have negative impact on VMM precision (Fig. \ref{bldep}b) and dynamic energy. Determining optimal value of $\Delta V_{\textrm{D}}$ and by how much CMOS transistor currents can be reduced without negatively impacting precision is important future research. The estimates, based on the implemented layout (with rather suboptimal value of $C$), $\Delta V_{\textrm{D}} = 0.2$ V, and $N=10$, show that the total energy is about 5.44 pJ for $10\times10$ VMM, or equivalently $38.6$ TOps/J, with the static energy contributes roughly $65\%$ of the total budget. The energy-efficiency significantly improves for larger VMMs, e.g. reaching $\sim 120$ TOps/J for $N$ = 100, due to reduced contribution of static energy component. It becomes even larger, potentially reaching $150$ TOps/J for $N$ = 1000, at which point it is completely dominated by dynamic energy related to charging/discharging external capacitor.   

\par The area breakdown by the circuit components was accurately evaluated from the layout (Fig. \ref{circuit}). Clearly, because of rather small implemented VMMs, the peripheral circuitry dominates, with one neuron block occupying $\sim$ 1.5 larger area than the whole $10\times20$ supercell array. However, with larger and more practical array sizes (e.g. $N > 200$), the area is completely dominated by the memory array and external capacitors, which occupy $\sim25\%$ and $\sim75\%$, respectively, of the total area for the conservative design. 

\par In some cases, e.g. convolutional layers in deep neural networks, the same matrix of weights (kernels) is utilized repeatedly to perform large number of multiplications. To increase density, VMM operations are performed using time-division-multiplexing scheme which necessitates storing temporal results and, for our approach, performing conversion between digital and time-domain representations.  Fortunately, the conversion circuitry for the proposed VMM can be very efficient due to digital nature of time-encoded input/output signals. We have designed such circuitry in which the input conversion is performed with a shared counter and a simple comparator-latch to create time-modulated pulse, while the pulse-encoded outputs are converted to digital signals with a help of shared counter and a multi-bit register. Figure \ref{efficiency} summarizes energy and area for a time-domain multiplier based on the conservative design, in particular showing that the overhead of the I/O conversion circuitry drops quickly and becomes negligible as VMM size increases.

\par With a more advanced design, which will require more detailed simulations, the external capacitor can be significantly scaled down or eliminated completely. (For example, the capacitive coupling can be efficiently suppressed by adding dummy input lines and using differential input signaling.) In this case, latency and energy will be limited by intrinsic parasitics of the memory cell array, and, e.g., can be below 2 ns and 1 fJ per operation, respectively, for 6-bit $1000\times1000$ VMM. Moreover, the energy and latency are expected to improve with scaling of CMOS technology (decrease proportionally to the feature size).

The proposed approach compares very favourably with previously reported work, such as current-based 180 nm FG/CMOS VMM with measured 5,670 GOps/J \cite{fg1}, current-based 180 nm CMOS 3-bit VMM with estimated 6,390 GOps/J \cite{indiveri}, switch-cap 40 nm CMOS 3-bit VMM with measured 7,700 GOps/J \cite{swcap}, memristive 22-nm 4-bit VMM with estimated 60,000 GOps/J \cite{memdpe2}, and ReRAM-based 14 nm 8-bit VMM with estimated 181.8 TOps/J \cite{reram}. Note that the most impressive reported energy efficiency numbers do not account for costly I/O conversion, specific to these designs.

\section{Summary}
\par We have proposed novel time-domain approach for performing vector-by-matrix computation and then presented the design methodology for implementing larger-scale circuit, based on the proposed multiplier, completely in time domain. As a case study, we have designed a simple multilayer perceptron network, which involves two layers of $10\times10$ four-quadrant vector-by-matrix multipliers, in 55-nm process with embedded NOR flash memory technology. In our performance study, we have focused on the detailed characterization of the most important factor limiting the precision, and then discussed key trade-offs and key performance metrics. The post-layout estimates for the conservative design which also includes the I/O circuitry to convert between digital and time-domain representation, show up to $>150$ TOps/J energy efficiency at $>5$ bit computing precision. A much higher energy efficiency, exceeding POps/J energy efficiency threshold, can be potentially achieved by using more aggressive CMOS technology and advanced design, though this opportunity requires more investigation.  

%%%%%%% -- PAPER CONTENT ENDS -- %%%%%%%%

%%%%%%%%% -- BIB STYLE AND FILE -- %%%%%%%%
\bibliographystyle{unsrt}
\bibliographystyle{ACM-Reference-Format}

\end{document}